\documentclass[twocolumn,letterpaper,nofootinbib,amsmath]{revtex4}

\pdfoutput=1
\usepackage{epsfig}
\usepackage{hyperref}

\newcommand{\lessthansimilarto}{\lower3pt\hbox{$\buildrel{<}\over{\sim}$}}
\newcommand{\greaterthansimilarto}{\lower3pt\hbox{$\buildrel{>}\over{\sim}$}}

\newcommand{\RR}{\hbox{$I$\kern-3.8pt $R$}}
\newcommand\dperp{{\partial}_\perp}

\begin{document}

\title{Covariant formulations of BSSN and the standard gauge}

\author{J.~David Brown}
\affiliation{Department of Physics, North Carolina State University,
Raleigh, NC 27695 USA}

\begin{abstract}
The BSSN and standard gauge equations are written in covariant form with respect 
to spatial coordinate transformations. The BSSN variables are defined as
tensors with no density weights. This allows us
to evolve a given set of initial data using two different coordinate systems and 
to relate the results using the familiar tensor transformation rules. 
Two variants of the covariant equations 
are considered. These differ from one another in the way that 
the determinant of the conformal metric is evolved. 
\end{abstract}
 
\maketitle

\section{Introduction}
The BSSN formulation of Einstein's equations \cite{Shibata:1995we,Baumgarte:1998te} 
is in widespread use in the numerical relativity 
community. These equations are most often used in conjunction with the ``standard gauge"
conditions, namely, 1+log slicing and the Gamma--driver shift. 

The BSSN variables include the conformal metric $g_{ab}$, conformal factor $\varphi$, 
and the trace and trace--free parts of the extrinsic curvature, $K$ and $A_{ab}$. 
They are defined in terms of the physical spatial metric $\hat g_{ab}$ 
and the physical extrinsic curvature $\hat K_{ab}$  by
\begin{subequations}\label{InverseDefinegK}
\begin{eqnarray}
	\hat g_{ab} & = & e^{4\varphi}g_{ab} \ ,\\
	\hat K_{ab} & = & e^{4\varphi}(A_{ab} + g_{ab} K/3) \ .
\end{eqnarray}
\end{subequations}
The conformal metric is chosen to have unit determinant, $g=1$, and $A_{ab}$ 
is trace--free.
A key ingredient of the BSSN formulation is the use of 
the ``conformal connection functions", defined by  
$\Gamma^a \equiv -\partial_b g^{ab}$. 

The 1+log slicing condition is an evolution equation for the lapse function 
$\alpha$ that takes the form \cite{Bona:1994dr}
\begin{equation}\label{onepluslog}
	\partial_t\alpha = \beta^c\partial_c\alpha  -2\alpha K \ ,
\end{equation}
where $\beta^a$ is the shift vector. 
The Gamma driver shift is defined by \cite{Alcubierre:2002kk}
\begin{subequations}\label{GammaDriver}
\begin{eqnarray}
	\partial_t \beta^a & = & \beta^c\partial_c\beta^a + \frac{3}{4} B^a \ ,\\
	\partial_t B^a & = & \beta^c\partial_c B^a + \partial_t\Gamma^a 
	- \beta^c\partial_c\Gamma^a  - \eta B^a \ ,
\end{eqnarray}
\end{subequations}
where $B^a$ is an auxiliary variable and $\eta$ is a constant. The term 
$\partial_t\Gamma^a$ in  Eq.~(\ref{GammaDriver}b) is 
replaced with the right--hand side of the equation of motion for $\Gamma^a$. 

The Gamma driver shift condition  is not generally covariant. In other words, 
Eqs.~(\ref{GammaDriver}) do not preserve their form under a time--independent 
transformation of the spatial coordinates. To see this, note that $\beta^a$ and 
$\partial_t \beta^a$ transform as contravariant vectors. The advection 
term $\beta^c\partial_c\beta^a$ in Eq.~(\ref{GammaDriver}a) 
does not transform as a contravariant vector. This spoils the covariance
of Eq.~(\ref{GammaDriver}a). 
Note, however, that the advection terms are not always included in the Gamma driver 
shift condition \cite{vanMeter:2006vi}. So for the moment let us ignore 
the terms with $\beta^c\partial_c$ acting on $\beta^a$, $B^a$, and $\Gamma^a$. 
Then Eq.~(\ref{GammaDriver}a) shows that $B^a$ and  $\partial_t B^a$ should
transform as contravariant vectors. However, the right--hand 
side of Eq.~(\ref{GammaDriver}b) depends on the connection functions 
$\Gamma^a$, which do not form a contravariant vector. 
Instead, $\Gamma^a$ obeys a rather complicated transformation rule determined 
from the following considerations. The conformal metric has unit determinant, 
$g = 1$. This equation is generally covariant under spatial coordinate 
transformations only 
if $g_{ab}$ is defined as a type $0\choose 2$ tensor density of weight $-2/3$. 
This makes $g$ a scalar, which is set equal to the scalar $1$. Therefore the  
conformal connection functions $\Gamma^a \equiv -\partial_b g^{ab}$  transform as 
the contraction of the derivative of a type $2\choose 0$ tensor density of weight $2/3$. This 
is a complicated transformation rule which I will not bother to write 
out in detail. The time derivative, $\partial_t\Gamma^a$, also satisfies this rule. 
We see that even if we ignore the advection terms, 
there is a mismatch in Eqs.~(\ref{GammaDriver}) in the way that the 
individual terms transform under time--independent changes of spatial coordinates. 

In addition to the conformal metric $g_{ab}$ and the conformal connection 
functions $\Gamma^a$, the BSSN variables include $\varphi$, $A_{ab}$, and $K$,
Because $g_{ab}$ carries 
a nonzero density weight, the conformal factor $\varphi$ must transform 
as the logarithm 
of a weight $1/6$ scalar density. The variable $A_{ab}$ is a trace--free 
type $0\choose 2$ tensor density of weight $-2/3$. 
The trace of the extrinsic curvature $K$ is a scalar. With these transformation 
rules, the BSSN equations and the 1+log slicing conditions are covariant. 

In section II, I discuss the issue of general covariance in more detail. This serves 
as further motivation for the subsequent analysis. In Section III, I rewrite the BSSN  
equations in terms of simple tensor variables with no density 
weights.  One of the key 
steps in the analysis is the introduction of a background connection, 
as suggested by Garfinkle, Gundlach, and Hilditch \cite{Garfinkle:2007yt}.
Another key 
step is to recognize that the condition $g=1$ should be replaced 
by an evolution equation for $g$ \cite{Brown:2005aq}. 

There are two natural 
choices for the evolution of $g$, which are presented in Section IV. One 
is the ``Lagrangian case" in which $\partial_t g =0$. Then $g$ is constant in 
time, equal to its initial value.  If the initial value of $g$ is unity 
and the background connection vanishes, these equations reduce to the 
traditional BSSN equations. These equations are not strongly hyperbolic 
unless the trace--free property of the variable $A_{ab}$ is actively enforced 
during the evolution. As an alternative, one 
can add a term proportional to $A$ to the evolution equation for the 
conformal metric. This yields a strongly hyperbolic system without the 
need to actively enforce the condition $A=0$. 

Another choice for the evolution of $g$ is
$\partial_t g = {\cal L}_\beta g$, where ${\cal L}$ is the Lie derivative. 
This is the ``Eulerian case". The BSSN equations in Eulerian form contain fewer 
terms than the traditional BSSN equations. 

In Section III the tensorial BSSN variables are used to rewrite the Gamma--driver 
shift condition in generally covariant form. If the initial value 
of $g$ is unity and the background connection vanishes, the covariant Gamma--driver
shift equations for the Lagrangian case 
reduce to the familiar Eqs.~(\ref{GammaDriver}). As an alternative,  the 
shift condition can be defined using Eqs.~(\ref{GammaDriver}) with 
the Eulerian evolution equation for $\partial_t\Gamma^a$. 

%%%%%%%%%%%%%%%%%%%%%
\section{Spatial covariance in numerical relativity}
Let me discuss the issue of spatial covariance in concrete terms. Let's say we are given 
a physical metric and extrinsic curvature, $\hat g_{ab}$ and $\hat K_{ab}$, that 
satisfy the Hamiltonian and momentum constraints.  
For simplicity, let us assume that these tensors 
$\hat g_{ab}$ and $\hat K_{ab}$ are expressed in terms of a single coordinate 
patch with ``Cartesian" coordinates $x$, $y$, and $z$. By calling the
coordinates Cartesian 
I mean  that each coordinate ranges over an interval of the real number line, with 
no periodic identification and no coordinate singularities. 
In this case we can transform to 
``spherical" coordinates $r$, $\theta$, and $\phi$  using the familiar relations 
$x = r\sin\theta \cos\phi$, $y = r\sin\theta \sin\phi$, and $z = r\cos\theta$. 
Let us denote the initial data in spherical coordinates by $\hat g_{ab}'$, 
$\hat K_{ab}'$. The initial data in these two coordinate systems are related by
\begin{subequations}\label{coordtransfhat}
\begin{eqnarray}
	\hat g_{ab}' & = & \frac{\partial x^c}{\partial {x'}^a} 
	\frac{\partial x^d}{\partial {x'}^b}\, \hat g_{cd} \ ,\\
	\hat K_{ab}' & = & \frac{\partial x^c}{\partial {x'}^a} 
	\frac{\partial x^d}{\partial {x'}^b}\, \hat K_{cd}  \ ,
\end{eqnarray}
\end{subequations}
where $x^a$ are Cartesian coordinates and ${x'}^a$ are spherical coordinates.

We want to evolve these data using the BSSN system. Starting with the 
physical metric and extrinsic curvature 
in Cartesian coordinates, $\hat g_{ab}$ and $\hat K_{ab}$, we apply the 
definitions (\ref{InverseDefinegK}) to obtain initial values for the 
BSSN variables $g_{ab}$, $\varphi$, $A_{ab}$, $K$, and $\Gamma^a$. Alternatively, we 
can start with the physical metric  
and extrinsic curvature in spherical coordinates, $\hat g_{ab}'$ 
and $\hat K_{ab}'$, and 
define the BSSN variables $g_{ab}'$, $\varphi'$, $A_{ab}'$, $K'$, and ${\Gamma'}^a$.
The primed and unprimed BSSN variables will be related by the 
coordinate transformation rules outlined in the introduction. For example,  
the relation for the conformal metric is 
\begin{equation}\label{coordtransf}
	g_{ab}' = \left| \frac{\partial x}{\partial x'} \right|^{-2/3} 
	\frac{\partial x^c}{\partial {x'}^a} 
	\frac{\partial x^d}{\partial {x'}^b}\,  g_{cd}  \ .
\end{equation}
The factor $|\partial x/\partial x'|$ is the Jacobian of the 
coordinate transformation. 

In order to evolve these data we must choose a lapse function $\alpha$ 
and shift vector $\beta^a$. 
Let us assume for definiteness that the lapse and shift are determined by 
evolution type equations and, for the moment, let us assume that these 
equations are spatially covariant.  For example, one might consider 
1+log slicing and a ``modified" Gamma driver condition obtained by 
replacing the term $\partial_t\Gamma^a -  \beta^c\partial_c\Gamma^a$ with, say, 
$\hat D_b \Sigma^{ab}$, and replacing the remaining 
spatial derivatives with covariant derivatives. 
(Here, $\hat D_b$ is the physical spatial covariant derivative and 
$\Sigma^{ab} \equiv (\hat D^a \beta^b - \alpha\hat K^{ab} )^{TF}$ 
is the distortion tensor. $TF$ denotes the trace--free part.)

These gauge conditions require us to specify initial values 
for the lapse, shift, and auxiliary variable. Let $\alpha$, $\beta^a$, and 
$B^a$ denote these initial values in Cartesian coordinates. The 
initial values in spherical coordinates are related by the 
familiar tensor transformation rules: For the scalar lapse we have $\alpha' = \alpha$, 
and for the contravariant vector shift, 
\begin{equation}
	{\beta'}^a = \frac{\partial {x'}^a}{\partial x^b} \beta^b \ .
\end{equation} 
The auxiliary variable $B^a$ also transforms as a contravariant vector. 

Now we're ready to evolve both the ``unprimed" Cartesian coordinate data and the ``primed" 
spherical coordinate data from the initial time $t_i$ to some final time $t_f$. How do 
the primed and unprimed BSSN variables compare at $t_f$? If, as we assumed above, 
the lapse and shift are determined by covariant relations, then the two sets of 
BSSN variables at time $t_f$ will be related by the same transformation rules 
that apply to the initial data. In particular, 
the two conformal metrics at time $t_f$ will be related by Eq.~(\ref{coordtransf}). 

This, of course, is a good situation. We would like to have the option of evolving 
our initial data using different spatial coordinate systems, and we would like to 
be able to compare the results. But the Gamma driver shift condition 
Eq.~(\ref{GammaDriver}) is not covariant. Thus, at times $t > t_i$, the 
shift vector obtained from 
the Cartesian coordinate evolution will differ {\em geometrically} from the shift vector
obtained from the spherical coordinate evolution. These two shift 
vectors will not be related by 
a coordinate transformation; rather, they will be geometrically distinct vector 
fields.\footnote{With the 1+log condition 
(\ref{onepluslog}), the slicing of spacetime does not depend on the 
shift vector or the coordinate system. If the advection term 
$\beta^c\partial_c \alpha$ is 
dropped, then a non--covariant shift condition will cause the slicing to 
depend on the coordinate system.}

Because the shift vector depends on the coordinate system, 
the BSSN variables at time $t_f$ will {\em not} be related by the 
coordinate transformation rules outlined above. 
In particular, the conformal metrics 
at time $t_f$ will not be related by Eq.~(\ref{coordtransf}). 
We can recombine the BSSN variables 
to form the physical metric and physical extrinsic curvature. The physical tensors
will not be related by the transformation 
(\ref{coordtransfhat}).

This shortcoming of the BSSN formulation with Gamma--driver shift can be fixed. 
This is done by rewriting the BSSN equations and the standard gauge conditions 
in terms of regular tensors with 
no density weights. The density weights are removed  by eliminating the requirement that 
the determinant of the conformal metric, $g$, should equal $1$ 
\cite{Brown:2005aq}. It is then necessary 
to specify an evolution equation for $g$. The simplest choice is to let $g$ be constant 
in time. Then $g$ is determined by its initial value, a weight $2$ scalar 
density called $\bar g$. The conformal metric $g_{ab}$ is then a type $0\choose 2$ 
tensor with no density weight. The next step is to define a variable
$\Lambda^a = g^{bc}(\Gamma_{bc}^a - \tilde \Gamma_{bc}^a)$ to take the place 
of the conformal connection functions \cite{Garfinkle:2007yt}. 
Here, $\Gamma_{bc}^a$ are the Christoffel symbols constructed from the 
conformal metric, and $\tilde\Gamma_{bc}^a$ is a background connection. 

The BSSN and standard gauge equations can be written in terms of the regular 
tensors $g_{ab}$, 
$\varphi$, $A_{ab}$, $K$, and $\Lambda^a$. These equations reduce to the 
traditional forms in use by numerical 
relativity groups when the density $\bar g$ is set to unity and the 
background connection is that of a flat metric in Cartesian 
coordinates: $\tilde\Gamma^a_{bc} = 0$. If we want to transform an initial data set from 
Cartesian to spherical coordinates, and preserve the tensor transformation rules under 
evolution, then we must transform $\bar g$ as a weight $2$ density and 
$\tilde\Gamma^a_{bc}$ as a connection. In particular, in spherical coordinates, $\bar g$ would 
no longer be $1$ and $\tilde\Gamma^a_{bc}$ would no longer vanish. 

%%%%%%%%%%%%%%%%%%%%%%%%%%%%%%%%%%%%%%%%%%%%%%%%%%%%%%%%%
\section{Tensor variables for BSSN}
In this section I derive the BSSN equations from scratch using only 
tensors with no density weights. Begin with the ``gdot--Kdot" form 
of the Einstein evolution equations:
\begin{subequations}\label{gdotKdot}
\begin{eqnarray}
	\dperp \hat g_{ab} & = & -2\alpha \hat K_{ab} \ ,\\
	\dperp \hat K_{ab} & = & \alpha \hat K \hat K_{ab} 
	- 2\alpha \hat K_{ac}\hat K^c_b \nonumber\\
	& & + \alpha \hat R_{ab} - \hat D_a \hat D_b \alpha \ .
\end{eqnarray}
\end{subequations}
The time derivative operator is defined by  $\dperp \equiv \partial_t - {\cal L}_\beta$ where 
${\cal L}_\beta$ is the Lie derivative along the shift. 
The Hamiltonian and momentum constraints are 
\begin{subequations}\label{HamMomConstraints}
\begin{eqnarray}
	{\cal H} & \equiv &  \hat K^2 - \hat K_{ab} \hat K^{ab} + \hat R \ ,\\
	{\cal M}_a & \equiv & \hat D_b ( \hat K^{b}_a - \hat K \delta^b_a) \ .
\end{eqnarray}
\end{subequations}
Indices on the momentum constraint and extrinsic curvature are raised and lowered with the physical metric. 

The BSSN variables are defined by Eqs~(\ref{InverseDefinegK}). However, we will 
not assume any restrictions on the determinant of $g_{ab}$ or the trace of $A_{ab}$. 
Then these definitions can be inverted to obtain 
\begin{subequations}\label{DefinegK}
\begin{eqnarray}
	g_{ab} & = & (\hat g/g)^{-1/3} \hat g_{ab} \ ,\\
	A_{ab} & = & (\hat g/g)^{-1/3} (\hat K_{ab} - \hat g_{ab} \hat K/3 + \hat g_{ab} A/3) \ ,\\
	\varphi & = & \frac{1}{12} \ln (\hat g/g) \ ,\\
	K & = & \hat K - A \ .
\end{eqnarray}
\end{subequations}
Note that   $A \equiv g^{ab} A_{ab}$ and $\hat K \equiv \hat g^{ab} \hat K_{ab}$. 
Defined in this way, the BSSN variables $g_{ab}$, $A_{ab}$, $\varphi$, 
and $K$ are tensors with no density weights. 

Now compute the time derivatives of Eqs.~(\ref{DefinegK}) using the gdot--Kdot equations 
(\ref{gdotKdot}), then use Eqs.~(\ref{InverseDefinegK}) to express 
the results in terms  of BSSN variables. This is a straightforward, although somewhat tedious
calculation. It is useful to note that 
$(\hat g/g)^{1/3} = e^{4\varphi}$, and for the last two terms in Eq.~(\ref{gdotKdot}b), 
\begin{subequations}\label{lasttwoterms}
\begin{eqnarray}
	\hat R_{ab} & = & R_{ab} - 2D_a D_b \varphi + 4D_a\varphi D_b\varphi 
		\nonumber\\
	& & - 2 g_{ab} (D^2\varphi + 2 D^c\varphi D_c\varphi) \ ,\\
	\hat D_a\hat D_b\alpha & = & D_a D_b \alpha - 4D_{(a}\alpha D_{b)}\varphi 
		\nonumber\\
	& & + 2g_{ab} D^c\alpha D_c\varphi  \ .
\end{eqnarray}
\end{subequations}
The result of this calculation is 
\begin{widetext} 
\begin{subequations}\label{evolutioneqns1}
\begin{eqnarray}
	\dperp  {g}_{ab}  & = & \frac{1}{3} g_{ab}\, \dperp\ln g -2{ \alpha} {  A}_{ab} 
			+ \frac{2}{3} \alpha g_{ab} A  \ ,\\
    \dperp {  A}_{ab} & = & \frac{1}{3} {  A}_{ab}\, \dperp \ln g + \frac{1}{3} g_{ab}\dperp A 
        -2{ \alpha} {  A}_{ac}{  A}^c_{b} 
       + { \alpha} {  A}_{ab} K   + \frac{1}{3} \alpha A (5A_{ab} - A g_{ab} - Kg_{ab} )\nonumber\\
       & &  + e^{-4{ \varphi}}  \left[ -2{ \alpha} D_a D_b { \varphi} 
	 + 4{ \alpha} D_a{ \varphi} D_b{ \varphi} + 4 D_{(a}{ \alpha} D_{b)} { \varphi} 
	 - D_a D_b{ \alpha} + { \alpha} R_{ab} \right]^{\rm TF} \ ,\\
    \dperp  { \varphi} 
        & = &  -\frac{1}{12} \, \dperp\ln g - \frac{1}{6} {  \alpha} (K + A) \ ,\\
	\dperp K & = & -\dperp A + \alpha(K + A)^2 + e^{-4\varphi}(\alpha R - 8\alpha D^a\varphi D_a\varphi
		- 8 \alpha D^2\varphi - D^2\alpha - 2D^a\alpha D_a\varphi)  \ .
\end{eqnarray}
\end{subequations}
\end{widetext}
The superscript ``$TF$" denotes the trace--free part of the expression in brackets. 

Now define 
\begin{subequations}\label{DefineGamma}
\begin{eqnarray}
	\Delta\Gamma^a_{bc} & \equiv & \Gamma^a_{bc} - \tilde\Gamma_{bc}^a \ ,\\
	\Delta\Gamma^a & \equiv & g^{bc}\Delta\Gamma^a_{bc}  \ ,
\end{eqnarray} 
\end{subequations}
where $\tilde\Gamma^a_{bc}$ is a background connection.  Although 
it is not necessary, it is convenient to think of the background connection 
as being built from a background metric, $\tilde g_{ab}$. 
I assume that the background connection is time independent. 
Note that $\Delta\Gamma^a$ 
is a contravariant vector.

From the definition of the Riemann tensor, we have the following identity: 
%% See RicciIdentity.nb
\begin{eqnarray}\label{RicciIdentity}
	  R_{ab} & \equiv & -\frac{1}{2} g^{cd} \tilde D_c \tilde D_d g_{ab} 
	+ g_{c(a}\tilde D_{b)}\Delta\Gamma^c \nonumber\\
	& & - g^{cd} g_{e(a} \tilde R_{b)cd}{}^e
	+ g^{de}\Delta\Gamma^c_{de} \Delta\Gamma_{(ab)c} \nonumber\\
      & & + g^{cd} \left( 2 \Delta\Gamma^e_{c(a} \Delta\Gamma_{b)ed} 
	+ \Delta\Gamma^e_{ac} \Delta\Gamma_{ebd} \right) \ .
\end{eqnarray}
Here, $\tilde D_a$ is the covariant derivative and $\tilde R^a{}_{bcd}$ is the Riemann 
tensor built from $\tilde\Gamma^a_{bc}$. 

The equations of motion (\ref{evolutioneqns1}) imply 
\begin{eqnarray}\label{EvolveGamma}
	\dperp (\Delta\Gamma^a) & = & 
		g^{bc} \tilde D_b \tilde D_c \beta^a - g^{bc} \tilde R^a{}_{bcd} \beta^d 
		-2A^{ab}\partial_b\alpha \nonumber\\
	& & - \frac{2\alpha}{\sqrt{g}} \tilde D_b ( \sqrt{g} A^{ab})
		- \frac{1}{3} \Delta\Gamma^a (\dperp\ln g + 2\alpha A) \nonumber\\
	& & 	- \frac{1}{6} g^{ab}\partial_b (\dperp\ln g - 4\alpha A)  \ .
\end{eqnarray}
Again, this is a straightforward, but tedious calculation. 
We now let $\Lambda^a$ denote a new BSSN variable which equals $\Delta\Gamma^a$ when the 
following constraint holds:
\begin{equation}\label{ConstraintC}
	{\cal C}^a \equiv \Lambda^a - \Delta\Gamma^a \ .
\end{equation}
$\Lambda^a$ is a contravariant vector. 

Next, we modify the equations of motion using the Hamiltonian and momentum constraints. 
The Hamiltonian constraint is ${\cal H} = 0$; from Eq.~(\ref{HamMomConstraints}a) 
we find 
\begin{eqnarray}\label{ConstraintH}
	{\cal H} & = & \frac{2}{3} (K + A)^2 + \frac{1}{3} A^2 - A_{ab} A^{ab} 
	\nonumber\\ 
	& & + e^{-4\varphi}(R - 8D^a\varphi D_a\varphi -8D^2\varphi) \ .
\end{eqnarray}
Now add $-\alpha {\cal H}$ to the right--hand side of Eq.~(\ref{evolutioneqns1}d).
The momentum constraint is ${\cal M}^a = 0$; from Eq.~(\ref{HamMomConstraints}b) 
we find 
\begin{eqnarray}\label{ConstraintM}
	{\cal M}^a & = & \frac{1}{\sqrt{g}}e^{-4\varphi} \tilde D_b ( \sqrt{g} A^{ab}) 
		+ 6e^{-4\varphi} (A^{ab} - A g^{ab}/3)\partial_b \varphi 
		\nonumber\\
	& & - e^{-4\varphi} g^{ab} \partial_b( 2K/3 + A) + e^{-4\varphi} A^{bc} 
		\Delta\Gamma^a_{bc} \ .
\end{eqnarray}
The equation of motion for $\Lambda^a$ is obtained by replacing 
$\dperp\Delta\Gamma^a$ on the left--hand side of Eq.~(\ref{EvolveGamma}) 
with $\dperp\Lambda^a$. We then add
$2\alpha e^{4\varphi}{\cal M}^a$ to the right--hand side of 
this equation. 

The analysis described above yields the following equations for the tensor 
BSSN variables:
%% See evolutioneqns2.nb
\begin{widetext}
\begin{subequations}\label{evolutioneqns2}
\begin{eqnarray}
	\dperp  {g}_{ab}  & = & \frac{1}{3} g_{ab}\, \dperp\ln g -2{ \alpha} {  A}_{ab} 
			+ \frac{2}{3} \alpha g_{ab} A  \ ,\\
    \dperp {  A}_{ab} & = & \frac{1}{3} {  A}_{ab}\, \dperp \ln g + \frac{1}{3} g_{ab}\dperp A 
        -2{ \alpha} {  A}_{ac}{  A}^c_{b} 
       + { \alpha} {  A}_{ab} K   + \frac{1}{3} \alpha A (5A_{ab} - A g_{ab} - Kg_{ab} )\nonumber\\
       & &  + e^{-4{ \varphi}}  \left[ -2{ \alpha} D_a D_b { \varphi} 
	 + 4{ \alpha} D_a{ \varphi} D_b{ \varphi} + 4 D_{(a}{ \alpha} D_{b)} { \varphi} 
	 - D_a D_b{ \alpha} + { \alpha} {\cal R}_{ab} \right]^{\rm TF} \ ,\\
    \dperp  { \varphi} 
        & = &  -\frac{1}{12} \, \dperp\ln g - \frac{1}{6} {  \alpha} (K + A) \ ,\\
	  \dperp  K & = &  -\dperp A + \frac{\alpha}{3}(K^2 + 2KA) 
	+ \alpha  A_{ab}A^{ab} 
        - e^{-4{ \varphi}} \left( D^2{ \alpha} + 2 D^a{  \alpha} D_a{ \varphi} \right)  
         \ ,\\	   
	\dperp \Lambda^a  & = &  g^{bc} {\tilde D}_b{\tilde D}_c \beta^a 
	- g^{bc} {\tilde R}^a{}_{bcd}\beta^d 
		-\frac{1}{3} \Delta\Gamma^a \,\dperp\ln g 
        - \frac{1}{6} g^{ab} \partial_b \dperp\ln g  \nonumber\\ 
	& & -2(A^{bc} - g^{bc}A/3)(\delta_b^a\partial_c \alpha 
	- 6\alpha \delta_b^a\partial_c\varphi - \alpha\Delta\Gamma^a_{bc}) 
	 - \frac{4}{3}\alpha  g^{ab} \partial_b K \ ,
\end{eqnarray}
\end{subequations}
where 
\begin{equation}\label{RicciDefinition}
	  {\cal R}_{ab}  \equiv  -\frac{1}{2} g^{cd} \tilde D_c \tilde D_d g_{ab} 
	+ g_{c(a}\tilde D_{b)}\Lambda^c - g^{cd} g_{e(a} \tilde R_{b)cd}{}^e  
	 + g^{de}\Delta\Gamma^c_{de} \Delta\Gamma_{(ab)c} 
	 + g^{cd} \left( 2 \Delta\Gamma^e_{c(a} \Delta\Gamma_{b)ed} 
	+ \Delta\Gamma^e_{ac} \Delta\Gamma_{ebd} \right) \ .
\end{equation}
\end{widetext}
In Eq.~(\ref{RicciDefinition}), ${\cal R}_{ab}$ is defined by using 
$\Lambda^a$ in place of $\Delta\Gamma^a$ in the identity
(\ref{RicciIdentity}).

The Eqs.~(\ref{evolutioneqns2}) are not complete evolution equations because
$\dperp g$ and $\dperp A$ appear on the right--hand sides. These equations are consistent 
in the sense that we can use them to compute $\dperp g$, and the result is 
an identity: $\dperp g = \dperp g$. Similarly, Eqs.~(\ref{evolutioneqns2}) yield 
an identity for $\dperp A$. 

Because the quantities $\dperp g$ and $\dperp A$ appear on the right--hand sides of 
Eqs.~(\ref{evolutioneqns2}), we must specify how $g$ and $A$ 
evolve. There are two natural choices for $g$, namely, $\partial_t g = 0$ 
and $\dperp g = 0$. In Ref.~\cite{Brown:2005aq,Brown:2007nt}, 
these were referred to as the {\em Lagrangian case} 
and the {\em Eulerian case}, respectively. For $A$, I will only consider 
the evolution equation $\partial_t  A  = 0$. These cases are described in detail 
in the next section. 

Using the tensorial variables define above, the standard gauge conditions are:
\begin{subequations}\label{StandardGauge}
\begin{eqnarray}
	\partial_t \alpha & = & \beta^c \tilde D_c \alpha - 2\alpha K  \ ,\\
	\partial_t \beta^a & = &  \beta^c \tilde D_c \beta^a  
	+ \frac{3}{4} B^a \ ,\\
	\partial_t B^a & = & \beta^c \tilde D_c B^a +  
	(\partial_t \Lambda^a - \beta^c \tilde D_c\Lambda^a )
	- \eta B^a \ .
\end{eqnarray}
\end{subequations}
Equation (\ref{StandardGauge}a)  is the 1+log slicing condition and 
Eqs.~(\ref{StandardGauge}b,c) are the Gamma--driver shift condition. 
The extra variable $B^a$ is a contravariant vector with no density weight.  
On the right--hand side of Eq.~(\ref{StandardGauge}c) the term $\partial_t\Lambda^a$ is 
eliminated using the BSSN equation of motion for $\Lambda^a$. 

%%%%%%%%%%%%%%%%%%%%%%%%%%%%%%
\section{Lagrangian and Eulerian cases}
For the Lagrangian case we have 
$\partial_t g = 0$ and $\partial_t A = 0$.  
For any choice of initial values, $g$ and $A$ will remain unchanged throughout 
the evolution. Let us call these initial values $\bar g$ and $\bar A$. Thus 
$\bar g$ is a time independent spatial scalar density of weight $+2$, and 
$\bar A$ is a time independent spatial scalar. 
The time independence of $g$ and $A$ imply $\dperp\ln g = -2D_a\beta^a$ and
$\dperp A = -\beta^a \partial_a A$. 

Since $g$ and $A$ equal their initial values for all time, we can replace 
$g$ with $\bar g$ and $A$ with $\bar A$ wherever they appear in Eqs.~(\ref{evolutioneqns2}).
Note that the covariant divergence of the shift depends on the 
spatial metric only through its determinant:
$D_a \beta^a = \partial_a ( \sqrt{\bar g} \beta^a )/\sqrt{\bar g}$. Since the determinant 
is constant in time, we can replace $D_a\beta^a$ with $\bar D_a\beta^a$, where 
$\bar D_a$ is the covariant derivative built from the initial conformal 
metric $\bar g_{ab}$. 
If we make the replacements $g \to \bar g$, $A \to\bar A$, and $D_a\beta^a \to \bar D_a\beta^a$ 
everywhere, we obtain the traditional BSSN equations written 
in covariant form. 

The traditional BSSN equations are not strongly hyperbolic unless the algebraic 
constraint $A = \bar A$ is continuously enforced \cite{Gundlach:2005ta,BSSNhyperbolicity}. 
(This is true for any choice of gauge 
conditions, not just the standard gauge.) In practice, the constraint 
$A = 0$ is imposed by making the replacement $A_{ab} \to A_{ab} - g_{ab} A/3$ 
after every sub--timestep in the numerical evolution. This prevents $A$ from developing 
a non--zero value due to numerical error. 

As an alternative, we can achieve strong hyperbolicity by leaving the term 
$2\alpha g_{ab} A/3$ in Eq.~(\ref{evolutioneqns2}a) alone \cite{BSSNhyperbolicity}. 
If we do this and also choose 
$\bar A = 0$, we find the following {\em Lagrangian BSSN equations}: 
\begin{widetext}
\begin{subequations}\label{ModifiedBSSN}
\begin{eqnarray}
	\dperp  {g}_{ab}  & = &  - \frac{2}{3} g_{ab} \bar D_c\beta^c -2{ \alpha} {  A}_{ab} 
			+ \frac{2}{3} \alpha g_{ab}  A  \ ,\\
    \dperp {A}_{ab} & = &  - \frac{2}{3} {  A}_{ab} \bar D_c\beta^c 
        -2{ \alpha} {  A}_{ac}{  A}^c_{b} 
       + { \alpha} {  A}_{ab} K   \nonumber\\
       & &  + e^{-4{ \varphi}}  \left[ -2{ \alpha} D_a D_b { \varphi} 
	 + 4{ \alpha} D_a{ \varphi} D_b{ \varphi} + 4 D_{(a}{ \alpha} D_{b)} { \varphi} 
	 - D_a D_b{ \alpha} + { \alpha} {\cal R}_{ab} \right]^{\rm TF} \ ,\\
    \dperp  { \varphi} 
        & = &  \frac{1}{6} \bar D_c\beta^c - \frac{1}{6} {  \alpha} K  \ ,\\
	  \dperp  K & = &  \frac{\alpha}{3}K^2 + \alpha A_{ab}A^{ab} 
        - e^{-4{ \varphi}} \left( D^2{ \alpha} + 2 D^a{  \alpha} D_a{ \varphi} \right)    \ ,\\
	\dperp \Lambda^a  & = &  {\cal C}^b \tilde D_b\beta^a 
	+ g^{bc} {\tilde D}_b{\tilde D}_c \beta^a - g^{bc} {\tilde R}^a{}_{bcd}\beta^d 
		+\frac{2}{3} \Delta\Gamma^a \bar D_c\beta^c 
        + \frac{1}{3} D^a \bar D_c\beta^c  \nonumber\\ 
	& & -2A^{bc}(\delta_b^a\partial_c \alpha 
		- 6\alpha\delta_b^a\partial_c\varphi - \alpha\Delta\Gamma^a_{bc}) 
	 - \frac{4}{3}\alpha g^{ab} \partial_b K \ .
\end{eqnarray}
\end{subequations}
\end{widetext}
${\cal R}_{ab}$ is given by Eq.~(\ref{RicciDefinition}).  
These equations with the standard gauge Eqs.(\ref{StandardGauge}) are strongly 
hyperbolic without 
the need to enforce $A = 0$ explicitly. If the term $2\alpha g_{ab} A/3$ is omitted from 
Eq.~(\ref{ModifiedBSSN}a), the result is the  {\em traditional BSSN equations}. 
The traditional BSSN equations coincide with the equations that are in 
widespread use in the numerical relativity community when the background 
connection vanishes, $\tilde\Gamma^a_{bc} = 0$, and the initial data satisfies 
$\bar g = 1$.  

It is a bit of an overstatement to say that the Lagrangian system (\ref{ModifiedBSSN}), or the 
traditional system with $A=0$ enforced, is strongly hyperbolic. As shown by Beyer 
and Sarbach \cite{Beyer:2004sv}, the traditional BSSN system plus standard 
gauge is strongly 
hyperbolic for $2\alpha \ne e^{4\varphi}$. 
The condition $2\alpha \ne e^{4\varphi}$ is likely violated on a surface of 
co--dimension one in black hole simulations, but in practice this 
does not seem to be a problem. 

Another natural choice for the evolution of $g$   is the 
{\em Eulerian case}, $\dperp g = 0$. Let us assume as before that 
$A$ is time independent, $\partial_t A = 0$. 
Let us replace $A$ with its initial value $\bar A = 0$ everywhere except in the 
$\dperp g_{ab}$  equation. Then the {\em Eulerian BSSN equations} are 
\begin{widetext}
\begin{subequations}\label{EulerianBSSN2}
\begin{eqnarray}
	\dperp  {g}_{ab}  & = &   -2{ \alpha} {  A}_{ab} 
			+ \frac{2}{3} \alpha g_{ab}  A  \ ,\\
    \dperp {A}_{ab} & = &  
        -2{ \alpha} {  A}_{ac}{  A}^c_{b} 
       + { \alpha} {  A}_{ab} K   %\nonumber\\ & & 
         + e^{-4{ \varphi}}  \left[ -2{ \alpha} D_a D_b { \varphi} 
	 + 4{ \alpha} D_a{ \varphi} D_b{ \varphi} + 4 D_{(a}{ \alpha} D_{b)} { \varphi} 
	 - D_a D_b{ \alpha} + { \alpha} {\cal R}_{ab} \right]^{\rm TF} \ ,\\
    \dperp  { \varphi} 
        & = &   - \frac{1}{6} {  \alpha} K  \ ,\\
	  \dperp  K & = &   \frac{\alpha}{3}K^2 
	+ \alpha A_{ab}A^{ab} 
        - e^{-4{ \varphi}} \left( D^2{ \alpha} 
	+ 2 D^a{  \alpha} D_a{ \varphi} \right)    \ ,\\
	\dperp \Lambda^a  & = &  {\cal C}^b \tilde D_b\beta^a 
	+ g^{bc} {\tilde D}_b{\tilde D}_c \beta^a - g^{bc} {\tilde R}^a{}_{bcd}\beta^d 
		   -2A^{bc} (\delta_b^a\partial_c \alpha 
		- 6\alpha\delta_b^a\partial_c\varphi - \alpha\Delta\Gamma^a_{bc}) 
	 - \frac{4}{3}\alpha g^{ab} \partial_b K \ .
\end{eqnarray}
\end{subequations}
\end{widetext}
Again, ${\cal R}_{ab}$ is given by Eq.~(\ref{RicciDefinition}). 
These equations with the standard gauge Eqs.~(\ref{StandardGauge}) 
are strongly hyperbolic for $2\alpha \ne e^{4\varphi}$ and do not require 
enforcement of the algebraic constraint $A = \bar A = 0$. They are more simple 
than both the Lagrangian equations (\ref{ModifiedBSSN}) and the traditional BSSN equations. 

Note that the Gamma driver shift (\ref{StandardGauge}b,c) depends on $\partial_t\Lambda^a$. 
This term is to be replaced with appropriate terms from the equation of motion for $\Lambda^a$. 
There are two possibilities. The term $\partial_t\Lambda^a$ can be defined using either 
the Lagrangian equation (\ref{ModifiedBSSN}e) or the Eulerian equation (\ref{EulerianBSSN2}e).  
If one wants the Gamma driver shift condition as it is currently defined 
in the numerical relativity community, then the Lagrangian equation 
should be used. This is the case even if 
one chooses to evolve the BSSN variables using the Eulerian 
Eqs.~(\ref{EulerianBSSN2}). The BSSN 
equations, either Lagrangian or Eulerian,  with the standard gauge that uses 
the Lagrangian equation for $\partial_t\Lambda^a$ to define the Gamma driver shift,  is
strongly hyperbolic for $2\alpha \ne e^{4\varphi}$. 

It would be interesting to investigate the properties of the shift condition 
defined by using the Eulerian equation (\ref{EulerianBSSN2}e) to eliminate 
$\partial_t\Lambda^a$ from Eqs.~(\ref{StandardGauge}). In this case the 
BSSN equations (either Lagrangian or Eulerian) plus gauge conditions are 
strongly hyperbolic for $8\alpha \ne 3 e^{4\varphi}$. A detailed analysis of 
hyperbolicity for these systems will be given elsewhere \cite{BSSNhyperbolicity}. 

%%%%%%%%%%%%%%%%%%%%%%%%%%%%%%%%%%%%%%%%
%%%%%%%%%%%%%%%%%%%%%%%%%%%%%%%%%%
\begin{acknowledgments}
I would like to thank Carl Meyer, Olivier Sarbach and Manuel Tiglio for helpful discussions. 
This work was supported by NSF grant PHY--0758116. 
\end{acknowledgments}

\bibliography{references}

\end{document}